# Effects due to unconventional pairing in transport through a normal metal-superconductor-normal metal hybrid junction

Ganesh C. Paul, Paramita Dutta, and Arijit Saha

# Effects Due to Unconventional Pairing in Transport Through a Normal Metal-Superconductor-Normal Metal Hybrid Junction

Ganesh C. Paul\*, Paramita Dutta, Arijit Saha

*Institute of Physics, Sachivalaya Marg, Bhubaneswar-751005, India.*
*Homi Bhabha National Institute, Training School Complex, Anushakti Nagar, Mumbai 400085, India*
*\*email: ganeshpaul@iopb.res.in*

**Abstract.** We explore transport properties of a normal metal-superconductor-normal metal (NSN) junction, where the superconducting region supports mixed singlet and chiral triplet pairings. We show that in the subgapped regime when the chiral triplet pairing amplitude dominates over that of the singlet, a resonance phenomena emerges out where all the quantum mechanical scattering probabilities acquire a value of 0.25 . At the resonance, crossed Andreev reflection mediating through such junction, acquires a zero energy peak. This reflects as a zero energy peak in the conductance as well in the topological phase when $\Delta_p > \Delta_s$.

## INTRODUCTION

Study of transport properties at the interface of normal metal-superconductor (NS) hybrid structures has been the topic of intense research interest during the last few decades. On the other hand, non-centrosymmetric superconductors (NCS) are examples of unconventional superconductors where the spin-singlet and triplet pairing mixing is present due to the presence time reversal symmetry but with broken inversion symmetry.

Very recently, transport signature of NS and superconductor-normal-superconductor (SNS) junction with mixed singlet and chiral triplet pairing has been reported by Burset et al. They obtained a zero-energy peak in the conductance in a NS junction when the triplet pairing dominates over the singlet part. However, NSN junction and the properties of CAR in the above context has never been studied so far. This motivated us to investigate transport phenomena through a NSN set up in which a one-dimensional (1D) nanowire (NW) is placed in close proximity to a superconductor which contains a pair potential of mixed singlet and chiral triplet type. The NW is attached to two normal metal (N) leads. We adopt Blonder-Tinkham-Klapwijk (BTK) formalism to calculate the quantum mechanical scattering amplitudes through the junction and conductance therein.

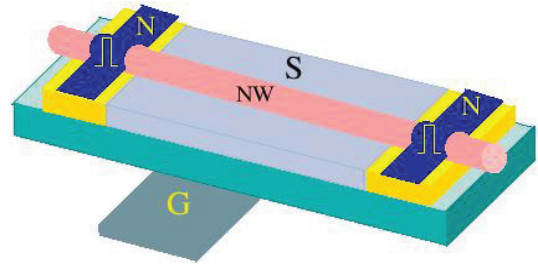

**FIGURE 1.** Schematic of our quasi one-dimensional NSN setup. The gate G is used for controlling the chemical potential inside the nanowire (NW). At each NS interface we consider an insulating barrier modelled by the δ-function potential.

## MODEL AND METHOD

In our schematic of set-up, we choose the x-axis along the direction of the NW. The two N-NW interfaces are located at x = 0 and x=L respectively. The NSN junction can be described by the Bogoliubov-deGennes (BdG) equations like,

$$H(\mathbf{k})\psi(\mathbf{k}) = \epsilon\psi(\mathbf{k})$$

where,

$$H(\mathbf{k}) = \begin{bmatrix} [E(\mathbf{k}) - \mu)]\hat{\sigma}_0 & \hat{\Delta}(\mathbf{k}) \\ \hat{\Delta}^\dagger(\mathbf{k}) & [\mu - E(-\mathbf{k})]\hat{\sigma}_0 \end{bmatrix}.$$

with the Nambu basis as,

$$\psi(\mathbf{k}) = [u_\uparrow(\mathbf{k}), u_\downarrow(\mathbf{k}), v_\uparrow(\mathbf{k}), v_\downarrow(\mathbf{k})]^T$$

$E(\mathbf{k})$ is the dispersion relation of the electronic excitation measured from the chemical potential μ. Generic form of pairing potential is given by,

$$\hat{\Delta}(\mathbf{k}) = i[\Delta_s(\mathbf{k})\hat{\sigma}_0 + \sum_{j=1}^{3} d_j(\mathbf{k})\hat{\sigma}_j]\hat{\sigma}_2 e^{i\phi}$$

The 4×4 Hamiltonian $H(\mathbf{k})$ can be decoupled into ↑↓ and ↓↑ channels. Pairing potentials corresponding to these channels take the form as,

$$\Delta_{1,2}(\theta) = [\Delta_s \pm \Delta_p e^{i\chi\theta}]\hat{\sigma}_2 e^{i\phi}$$

We write down the wave functions in three regions *i.e* left normal metal, NW and right normal metal and using boundary conditions at the two interfaces x = 0 and x=L, we find all the four scattering amplitudes $r_\sigma$, $r_{h\sigma}$, $t_{e\sigma}$ and $t_{h\sigma}$ which represent normal reflection (R), Andreev reflection (AR), cotunneling (CT) and crossed Andreev reflection (CAR), respectively.

At zero temperature, conductance for a particular incident electron energy and a chiral angle θ can be found by taking contributions from both the spin channel σ = 1 and 2 using the following relation,

$$G(\epsilon, \theta) = \frac{e^2}{h}\sum_\sigma \left(|t_{e\sigma}|^2 - |t_{h\sigma}|^2\right)$$

We normalize the conductance by the normal state conductance $G_0 = (2e^2/h) D(\theta)$, where, $D(\theta) = 4\cos^2\theta/(Z^2 + 4\cos^2\theta)$. As θ represents the relative orientation between the triplet and singlet components of the pairing potential, its range can be $[-\pi/2, \pi/2]$ with respect to the direction of the incoming electron. Therefore, the angle-averaged conductance can be obtained after integrating $G(e, \theta)$ over θ as,

$$\tilde{G}(\epsilon) = \int_{-\pi/2}^{\pi/2} G(\epsilon, \theta)\cos\theta \, d\theta$$

## RESULT AND DISCUSSION

Depending on the relative strength of triplet to singlet phase of the superconducting pairing potential we consider three different regimes of interest: $\Delta_p < \Delta_s$, $\Delta_p = \Delta_s$ and $\Delta_p > \Delta_s$. Although we present all our numerical results only for the regime $\Delta_p > \Delta_s$ which is in DIII topological class for θ = 0. Also, we show $R_{e\sigma}$, $R_{h\sigma}$, $T_{e\sigma}$ and $T_{h\sigma}$ (R, AR, CT and CAR probabilities respectively) as functions of different parameters of the system only for σ = 1 without loss of generality and hence we use the notation $R_e$, $R_h$, $T_e$ and $T_h$ respectively throughout our results. Values of parameters are chosen as Z = 2, φ = 0, e = 1, h = 1, μ = 1 and U = 15 (for the superconducting region). We use the unit where $\Delta_0$ = 1. Length of the superconducting region and energy of the incident electron are normalized by the superconducting coherence length (ξ) and amplitude of the pair potential $\Delta_0$.

In Fig. 2 we present all the four possible quantum mechanical scattering probabilities $R_e$, $R_h$, $T_e$ and $T_h$ as a function of the length (L) of the superconductor for $\Delta_p > \Delta_s$ regime, with incident electron energy e = 0, where panel (a) and (b) in Fig. 2 correspond to θ = 0 and θ = π/4 respectively. It is evident from Fig. 2(a) that for θ = 0, AR dominates over all other scattering processes except for very small values of L. To illustrate this, we show $R_e$, $R_h$, $T_e$ and $T_h$ in the inset of Fig. 2(a), for small values of L (L<<ξ). Note that all the scattering probabilities are almost identical to each other for L < 0.03ξ i.e. they occur with almost equal probability of value = 0.25 which is in sharp contrast to the $\Delta_p < \Delta_s$ regime where $T_h$ (CAR) is vanishingly small. On the other hand, for L > 0.07ξ, all scattering processes except AR become vanishingly small.

To investigate whether the above mentioned resonance phenomena persists for other values of θ, we show the behavior of $R_e$, $R_h$, $T_e$ and $T_h$ as a function of L in Fig. 2(b) for θ = π/4. Note that the scattering probabilities no longer remain equal in any regime. Instead, all of them become oscillatory with respect to L/ξ. With the enhancement of the length of the superconducting region, normal reflection dominates over all other processes while CT and CAR become vanishingly small in magnitude. These periodic variation with L can be manifested as the interference between the electron and hole wave-functions inside the superconducting region..

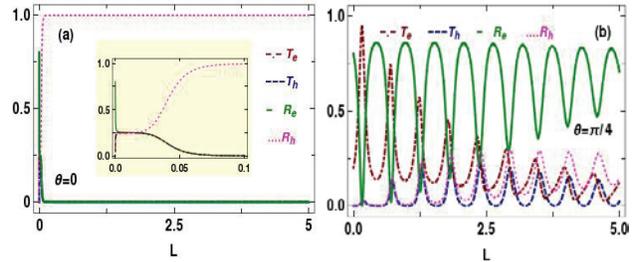

**FIGURE 2.** Quantum-mechanical scattering probabilities are plotted as a function of the length (L) of the superconducting region. In the inset of panel (a), the behavior of the scattering probabilities are shown when L<<ξ. The value of the other parameters are chosen to be $\Delta_p$ = 0.75, $\Delta_s$ = 0.25 .

In Fig. 3 we show the variation of $R_e$, $R_h$, $T_e$ and $T_h$ with incident electron energy . Here panel (a) and (b) correspond to $\theta = 0$ and $\theta = \pi/4$ respectively. The inset of Fig. 3(a) illustrates that AR exhibits a very sharp Zero Energy Peak (ZEP). This zero energy peak appears due to the presence of a zero energy Andreev bound state in the topological phase, $\Delta_p > \Delta_s$ . A very small deviation of energy from zero leads to a resonance where all the scattering processes achive a value ~ 0.25 as can be seen from inset of panel (a) of Fig. 3. On the other hand the probability for $R_e$ dominates for energy values other than zero. On the contrary, for $\theta = \pi/4$, the ZEP no longer exists as shown in Fig. 3(b). There are two resonance points symmetric around $\approx \pm 0.53\Delta_0$ in the subgapped regime where all the scattering processes acquire a value 0.25. Both AR and CAR have sharp peaks whereas the other two processes (R and CT) have dips at those points (see the inset of Fig. 3(b)). The change in the value of $\theta$ leads to the modification in the Andreev bound states present in the superconducting region resulting in resonance phenomena at different energies other than zero.

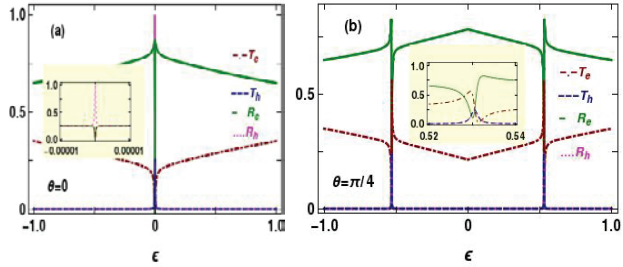

**FIGURE 3.** The behavior of scattering probabilities are shown as a function of the incident electron energy ($\epsilon$) in the subgapped regime. We choose $L = 0.01\xi$ .

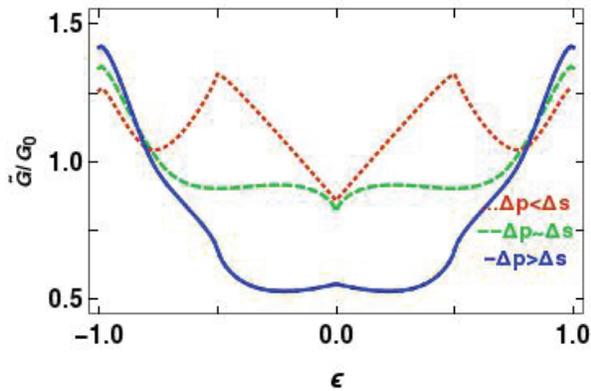

**FIGURE 4.** The behavior of normalized conductance ($G/G_0$) is shown as a function of the energy ($\epsilon$) of the incident electron for three different regime of mixed pairing potential ($\Delta_p < \Delta_s$ , $\Delta_p = \Delta_s$ and $\Delta_p > \Delta_s$ ). Here we choose $L = 1.2\xi$.

Normalized conductance ($G/G_0$) is presented in Fig. 4 where, we have averaged over all possible orientations between the singlet and triplet pair potentials. Note that conductance behavior is non-monotonic. There are peaks at $\epsilon = \pm\Delta_0$ for all the regimes and these peaks correspond to the density of states at the two boundaries of the superconducting gap. In the scattering probability profiles we obtain ZEP for AR process in the regime $\Delta_p > \Delta_s$, as a consequence of which there is a finite contribution in conductance at zero energy after averaging over all possible $\theta$s. The finite conductance at other values of $\epsilon$ can be attributed to the finite scattering probabilities at other values of theta.

## CONCLUSION

In summary, we have investigated the scattering probabilities and conductance phenomenon through a NSN junction where the superconductor is characterized by a mixture of both the spin-singlet and spin-triplet pairings. The main feature we obtain in this geometry is the appearance of a zero energy resonance. At the resonance, probability for all the four possible scattering processes (reflection, AR, CT and CAR) acquire same magnitude (1/4) when chiral triplet pairing amplitude dominates over the singlet one. The angle averaged conductance also exhibits a finite zero energy peak in this regime.

## REFERENCES

1. P. Burset, F. Keidel, Y. Tanaka, N. Nagaosa, and B. Trauzettel, Phy. Rev. B 90, 085438 (2014).
2. G. Blonder, M. Tinkham, and T. M. Klapwijk, Phy. Rev. B. 25, 4515 (1982).
3. M. Sigrist and K. Ueda, Rev. Mod. Phys. 63, 239 (1991).
4. A. Altland and M.R. Zirnbauer, Phy. Rev. B. 55,1142 (1997).